\documentclass[12pt,a4paper]{article}

\usepackage{authblk}
\usepackage{amsmath,amssymb}
\usepackage{graphicx}
\usepackage{amsfonts}
\usepackage{epsfig}
\usepackage{times,pstricks,pst-node}

\usepackage{pstricks,pst-node,pst-text,pst-3d}

\newcommand{\tr}{\operatorname{Tr}}
\newcommand{\llg}{\operatorname{ln}}

\newcommand{\ket}[1]{\left|#1\right\rangle}      
\newcommand{\bra}[1]{\left\langle #1\right|}     
\newcommand{\eq}{\begin{equation}}
\newcommand{\en}{\end{equation}}
\newcommand{\bear}{\begin{eqnarray}}
\newcommand{\ear}{\end{eqnarray}}

\title{Thermodynamic limit of the six-vertex model with reflecting end}

\author{G.A.P. Ribeiro\footnote{E-mail: pavan@df.ufscar.br; On leave of absence from Departamento de F\'isica, Universidade Federal de S\~ao Carlos, PO Box 676, 13565-905, S\~ao Carlos, SP, Brazil. } \ and V.E. Korepin \footnote{E-mail: korepin@gmail.com}}
\affil{C.N. Yang Institute for Theoretical Physics, \\
State University of New York at Stony Brook, \\
Stony Brook, NY 11794-3840, USA}

\begin{document}
\maketitle

\begin{abstract}
We study the thermodynamic limit of the six-vertex model with domain wall boundary and reflecting end. We evaluated  the partition function explicitly  in  special cases. We calculated  the homogeneous limit of the Tsuchiya determinant formula for the partition function. We evaluated  the thermodynamic limit and obtain the free energy of the six-vertex model with reflecting end. We  determined the free energy  in the disordered regime.
\end{abstract}

\thispagestyle{empty}

\newpage

\pagestyle{plain}
\pagenumbering{arabic}

\section{Introduction}

The six-vertex model with periodic boundary condition has been largely studied by Bethe ansatz techniques \cite{BAXTER,BOOK}. In the computation of scalar products of the Bethe states in the context of the quantum inverse scattering method \cite{FADDEEV}, it was introduced the six-vertex model with domain wall boundary condition (DWBC) \cite{KOREPIN1982}. In this case of fixed boundary condition, the partition function was given in terms of a determinant expression \cite{KOREPIN1992}. This determinant expression allowed for connections with enumerative combinatorics problems, e.g the proof of the number of alternating sign matrices \cite{KUPERBERG1996}.

Moreover, the partition function of the six-vertex model with domain wall boundary was studied in the thermodynamic limit \cite{KOREPIN2000,ZINNJUSTIN}. The results for the thermodynamic quantities like free energy and entropy were surprisingly different from the case of periodic boundary condition\cite{BAXTER}. Therefore, the role of boundary condition for the six-vertex vertex model becomes fundamental even in the thermodynamic limit.

Nevertheless, one can rise the question about the value of the thermodynamic quantities of the six-vertex model constrained by different fixed boundary conditions. In order to investigate the dependence of the physical quantities with the boundary conditions in the thermodynamic limit, we chose to consider another instance of integrable boundary. This addresses to the case where on the vertical direction one still has domain wall like boundary, however on the horizontal direction one has a reflecting end  \cite{TSUCHIYA}. Our main goal is to compute the free energy in the thermodynamic limit of the partition function with domain wall boundary condition and reflecting end. This is another non-trivial example where the boundary condition plays an import role.

The outline of the article is as follows. In section \ref{sixvertex}, we describe the six-vertex model and its boundaries conditions. In section \ref{det}, we discuss the partition function representation and its properties needed in this work. In section \ref{thermo}, we obtain the free energy in the thermodynamic limit. In section \ref{ent}, we compute the entropy in the disordered regime. Our conclusions are given on section \ref{CONCLUSION}.

\section{The six-vertex model}\label{sixvertex}

In this section, we describe the six-vertex model and its integrable boundaries conditions.

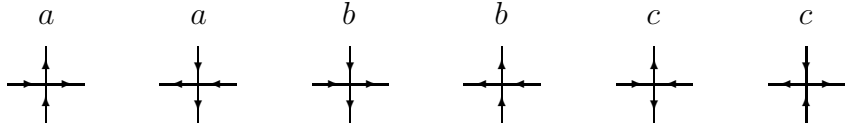
\begin{figure}[h]
\unitlength=1.0mm
\begin{center}
\begin{picture}(120,20)
\put(4,13){$a$}
\put(5,0){\line(0,1){10}}
\put(5,7.5){\vector(0,1){1}}
\put(5,2.5){\vector(0,1){1}}
\put(0,5){\line(1,0){10}}
\put(2.5,5){\vector(1,0){1}}
\put(7.5,5){\vector(1,0){1}}
\put(24,13){$a$}
\put(25,0){\line(0,1){10}}
\put(25,7.5){\vector(0,-1){1}}
\put(25,2.5){\vector(0,-1){1}}
\put(20,5){\line(1,0){10}}
\put(22.5,5){\vector(-1,0){1}}
\put(27.5,5){\vector(-1,0){1}}
\put(44,13){$b$}
\put(45,0){\line(0,1){10}}
\put(45,2.5){\vector(0,-1){1}}
\put(45,7.5){\vector(0,-1){1}}
\put(40,5){\line(1,0){10}}
\put(42.5,5){\vector(1,0){1}}
\put(47.5,5){\vector(1,0){1}}
\put(64,13){$b$}
\put(65,0){\line(0,1){10}}
\put(65,2.5){\vector(0,1){1}}
\put(65,7.5){\vector(0,1){1}}
\put(60,5){\line(1,0){10}}
\put(62.5,5){\vector(-1,0){1}}
\put(67.5,5){\vector(-1,0){1}}

\put(84,13){$c$}
\put(85,0){\line(0,1){10}}
\put(85,7.5){\vector(0,1){1}}
\put(85,2.5){\vector(0,-1){1}}
\put(80,5){\line(1,0){10}}
\put(82.5,5){\vector(1,0){1}}
\put(87.5,5){\vector(-1,0){1}}
\put(104,13){$c$}
\put(105,0){\line(0,1){10}}
\put(105,7.5){\vector(0,-1){1}}
\put(105,2.5){\vector(0,1){1}}
\put(100,5){\line(1,0){10}}
\put(102.5,5){\vector(-1,0){1}}
\put(107.5,5){\vector(1,0){1}}
\end{picture}
\caption{The Boltzmann weights of the six-vertex model.}
\label{6vert}
\end{center}
\end{figure}

The basic object containing the statistical weights of the six-vertex model is the $R$-matrix, which is given by\cite{BAXTER,BOOK}
\eq
R(\lambda)=\left(\begin{array}{cccc}
       a(\lambda) & 0 & 0 & 0 \\
       0 & c(\lambda)& b(\lambda) & 0  \\
       0 & b(\lambda) & c(\lambda) & 0 \\
       0 & 0 & 0 & a(\lambda)
\end{array} \right),
\en
where $a(\lambda), b(\lambda)$ and $c(\lambda)$ are the Boltzmann weights, which are associated to the different vertices configurations of the six-vertex model (see Figure \ref{6vert}).

The above $R$-matrix is a solution of the Yang-Baxter equation, 
\eq
R_{12}(\lambda-\mu)R_{23}(\lambda) R_{12}(\mu) =R_{23}(\mu)R_{12}(\lambda) R_{23}(\lambda-\mu),
\label{yangbaxter}
\en
which constraints the Boltzmann weights of the six-vertex such that,
\eq
\Delta=\frac{a^2+b^2-c^2}{2 a b},
\en
for any value of the spectral parameter. 

The Yang-Baxter equation (\ref{yangbaxter}) provides the commutativity property of the transfer matrix $T(\lambda)=\tr_{\cal A}\left[{\cal T}_{\cal A}(\lambda) \right]$, where the monodromy matrix is ${\cal T}_{\cal A}(\lambda)={\cal L}_{{\cal A}N}(\lambda-\mu_N)\cdots {\cal L}_{{\cal A}1}(\lambda-\mu_1)$, ${\cal L}_{12}(\lambda)=P_{12}R_{12}(\lambda)$ and $P_{12}$ is the permutation operator. 

The transfer matrix $T(\lambda)$ when multiplied successively builds up the partition function of a bidimensional classical vertex model with periodic boundary condition. The case of periodic boundary condition was extensively studied \cite{BAXTER}.

Within the quantum inverse scattering method one is able to diagonalize the transfer matrix and the quantum Hamiltonian simultaneously \cite{FADDEEV,BOOK}. One of the main ingredients is the algebraic relation among the monodromy matrix ${\cal T}_{\cal A}(\lambda)$ elements,
\eq
{\cal T}_{\cal A}(\lambda)=\left(\begin{array}{cc}
                            A(\lambda) & B(\lambda) \\
                            C(\lambda) & D(\lambda) 
                           \end{array}\right).
\en

As a result, the ansatz for the eigenstates can be written\cite{FADDEEV},
\eq
\ket{\psi}_N=B(\lambda_N) \cdots B(\lambda_2) B(\lambda_1)\ket{\Uparrow},
\en
where $\ket{\Uparrow}=\ket{\uparrow\cdots\uparrow}$ is the reference state taken as the ferromagnetic state. This ansatz provides the eigenvalues of the transfer matrix and consequently it determines the partition function with periodic boundary condition.

\subsection{Domain wall boundary condition}

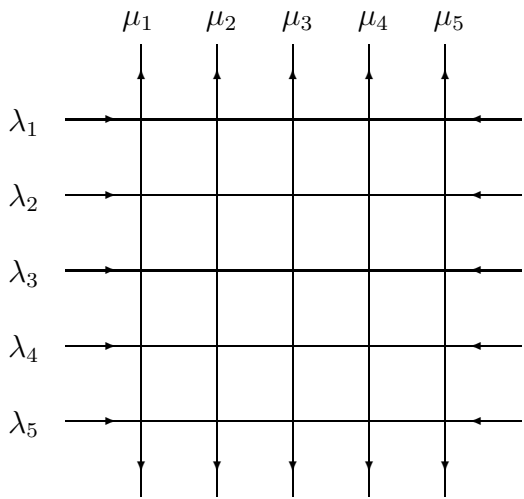
\begin{figure}[h]
\unitlength=0.5mm
\begin{center}
\begin{picture}(130,130)(-30,-10)

\multiput(-20,0)(0,20){5}{\line(1,0){120}}
\multiput(0,-20)(20,0){5}{\line(0,1){120}}
\multiput(-7.5,0)(0,20){5}{\vector(1,0){1}}
\multiput(87.5,0)(0,20){5}{\vector(-1,0){1}}
\multiput(0,-12.5)(20,0){5}{\vector(0,-1){1}}
\multiput(0,92.5)(20,0){5}{\vector(0,1){1}}

\put(-35,-3){$\lambda_5$}
\put(-35,17){$\lambda_4$}
\put(-35,37){$\lambda_3$}
\put(-35,57){$\lambda_2$}
\put(-35,77){$\lambda_1$}

\put(-5,105){$\mu_1$}
\put(17,105){$\mu_2$}
\put(37,105){$\mu_3$}
\put(57,105){$\mu_4$}
\put(77,105){$\mu_5$}

\end{picture}
\end{center}
\caption{The partition function $Z_N^{DWBC}$ for $N=5$ of the six-vertex model with domain wall boundary condition.}
\label{fig-DWBC}
\end{figure}

In the computation of scalar products of the above Bethe states it appears another distinguished partition function with fixed boundary conditions (see Figure \ref{fig-DWBC}), the so called domain wall boundary condition (DWBC)\cite{KOREPIN1982}
\eq
Z_N^{DWBC}(\{\lambda\},\{\mu\})=\bra{\Downarrow}B(\lambda_N)\cdots B(\lambda_2) B(\lambda_1)\ket{\Uparrow}.
\en

The above partition function can be cast in a determinant form \cite{KOREPIN1992}. This determinant formula pave the way to the understanding of the thermodynamic limit of the six-vertex model with DWBC. The results for the thermodynamic quantities, like free energy and entropy, were surprisingly different in comparison with usual periodic boundary\cite{KOREPIN2000,ZINNJUSTIN}. These results as well as its finite size corrections were rigorously proven \cite{BLEHER}.

\subsection{Reflecting end boundary condition}

Another instance of integrable boundary condition is the case of open boundary condition devised by Sklyanin \cite{SKYLIANIN}. In this case, the notion of integrability was extended so that the $R$-matrix continues describing the bulk dynamics and a new set of matrices, the $K$-matrices, represent the interaction at ends. This is a consequence of the reflection equation, which reads \cite{SKYLIANIN}, 
\eq
R_{12}(\lambda-\mu)K_{1}(\lambda)R_{21}(\lambda+\mu)K_{2}(\mu) =K_{2}(\mu) R_{12}(\lambda+\mu) K_{1}(\lambda) R_{21}(\lambda-\mu).
\label{refle}
\en

\begin{figure}[h]
\unitlength=1.mm
\begin{center}
\begin{picture}(50,15)
\put(12.5,4){$k_{+}$}
\put(0,0){\line(1,0){5}}
\put(3,0){\vector(1,0){1}}
\put(0,10){\line(1,0){5}}
\put(3,10){\vector(-1,0){1}}
\put(5,5){\oval(10, 10)[r]}
\put(52.5,4){$k_{-}$}
\put(40,0){\line(1,0){5}}
\put(43,0){\vector(-1,0){1}}
\put(40,10){\line(1,0){5}}
\put(43,10){\vector(1,0){1}}
\put(45,5){\oval(10, 10)[r]}

\end{picture}
\caption{The weights of the reflection end.}
\label{Krefle}
\end{center}
\end{figure}
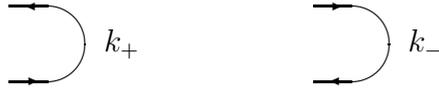

In the case of open boundary conditions, the transfer matrix can be written as
\eq
t(\lambda)=\tr_{\cal A}\left[\widetilde{K}_{\cal A}(\lambda){\cal T}_{\cal A}(\lambda)K_{\cal A}(\lambda) \widetilde{\cal T}_{\cal A}(\lambda)\right],
\label{opent}
\en
\eq
\widetilde{\cal T}_{\cal A}(\lambda)= {\cal L}_{{\cal A}1}(\lambda+\mu_1)\cdots {\cal L}_{{\cal A}N}(\lambda+\mu_N) \propto \left[{\cal T}_{\cal A}(-\lambda)\right]^{-1},
\en
where in the simplest case, the $K$-matrix  is a diagonal matrix (see Figure \ref{Krefle}),
\eq
K(\lambda)=\left(\begin{array}{cc}
            k_{+}(\lambda) & 0 \\
            0 & k_{-}(\lambda)
           \end{array}\right).
\en
The $\widetilde{K}$-matrix is related with $K(\lambda)$ due to some special symmetries\cite{SKYLIANIN}.

The Sklyanin's monodromy matrix is given by
\eq
U(\lambda)={\cal T}_{\cal A}(\lambda)K_{\cal A}(\lambda) \widetilde{\cal T}_{\cal A}(\lambda)=\left(\begin{array}{cc}
                            {\cal A}(\lambda) & {\cal B}(\lambda) \\
                            {\cal C}(\lambda) & {\cal D}(\lambda) 
                           \end{array}\right).
\label{openT}
\en
Thanks to the reflection equation (\ref{refle}), one has an additional algebra among these new monodromy matrix elements, which is called reflection algebra.

This allows us to define a new ansatz for the eigenstates of (\ref{opent}) \cite{SKYLIANIN}
\eq
\ket{\phi}_N={\cal B}(\lambda_N) \cdots {\cal B}(\lambda_2) {\cal B}(\lambda_1)\ket{\Uparrow}.
\label{openBS}
\en

\begin{figure}[h]
\unitlength=0.4mm
\begin{center}
\begin{picture}(80,140)(-30,-10)

\multiput(-20,0)(0,20){6}{\line(1,0){70}}
\multiput(0,-20)(20,0){3}{\line(0,1){140}}
\multiput(50,10)(0,40){3}{\oval(20, 20)[r]}
\multiput(-10,0)(0,20){6}{\vector(1,0){1}}
\multiput(0,-12.5)(20,0){3}{\vector(0,-1){1}}
\multiput(0,112.5)(20,0){3}{\vector(0,1){1}}

\put(-35,-3){$\lambda_3$}
\put(-40,17){$-\lambda_3$}
\put(-35,37){$\lambda_2$}
\put(-40,57){$-\lambda_2$}
\put(-35,77){$\lambda_1$}
\put(-40,97){$-\lambda_1$}

\put(-5,125){$\mu_1$}
\put(17,125){$\mu_2$}
\put(37,125){$\mu_3$}

\end{picture}
\end{center}
\caption{The partition function $Z_N$ for $N=3$ of the six-vertex model with reflecting end.}
\label{fig1}
\end{figure}
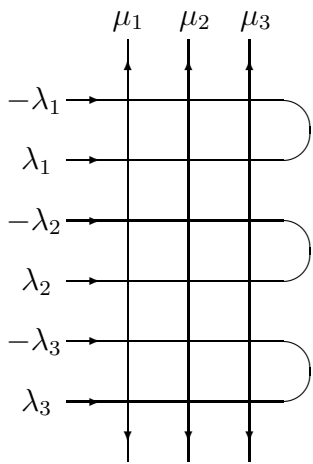

Again, the computation of the scalar products of these states (\ref{openBS}) leads naturally to a third distinguished partition function for the six-vertex model, which is due to Tsuchiya \cite{TSUCHIYA}. On the vertical direction, one still has domain wall like boundary. However, on the horizontal direction one has a reflecting end (see Figure \ref{fig1}), which has also been called $U$-turn boundary,
\eq
Z_N(\{\lambda\},\{\mu\})=\bra{\Downarrow}{\cal B}(\lambda_N)\cdots {\cal B}(\lambda_2) {\cal B}(\lambda_1)\ket{\Uparrow}.
\label{ZUturn}
\en

The determinant formula of $Z_N(\{\lambda\},\{\mu\})$ was also given in \cite{TSUCHIYA} and inspired some development in the combinatorics related to the number of vertically symmetric alternating sign matrices \cite{KUPERBERG}. Recently this partition function was also shown to be determined by functional relations and was expressed as multiple-contour integral \cite{GALLEAS}. The study of boundary correlations for the case of domain wall boundary conditions \cite{COLOMO} was also extended to the case of reflecting end boundary \cite{MOTEGI}.

In the context of open spin chains, the Tsuchiya determinant for the partition function yields in the surface free energy of the spin chain \cite{GOHMANN}. On the other hand, from the perspective classical vertex model, the above partition function describes the six-vertex model on a $N \times 2 N$ lattice with fixed boundary conditions (Figure \ref{fig1}).

\section{Determinant representation and Toda chain hierarchy}\label{det}

In this section, we define the determinant expression of the inhomogeneous six-vertex model with domain wall and reflecting end \cite{TSUCHIYA}. Then we discuss its homogeneous limit. We also note that this peculiar determinant formula satisfies the bidimensional Toda equation\cite{DARBAUX}. This equation will have an import role in the determination of the thermodynamic limit.

One assumes the usual parametrization for the Boltzmann weights\cite{BAXTER},
\eq
a(\lambda)=\sin(\gamma-\lambda), \qquad b(\lambda)=\sin(\gamma+\lambda), \qquad c(\lambda)=\sin(2\gamma),
\en
where $0<\gamma<\pi/2$ and $\Delta=-\cos(2\gamma)$ in the regime $-1<\Delta<1$. 

It is important to note that according to the definition of the Sklyanin's monodromy matrix (\ref{openT}), the following combination of the above weights appears in the partition function (\ref{ZUturn}),
\eq
a_{\pm}=a(\lambda\pm\mu)=\sin(\gamma-(\lambda\pm\mu)),\qquad b_{\pm}=b(\lambda\pm\mu)=\sin(\gamma+\lambda\pm\mu).
\en
This means that according to the Figure \ref{fig1}, in each pair of $U$-turn connected horizontal lines, the top horizontal lines contain Boltzmann weights $a_+,b_+,c$ and the lower ones contain the weights $a_-,b_-,c$.

Besides that, the reflection equation (\ref{refle}) determines the $K$-matrix elements as
\eq
k_{+}(\lambda)=\frac{\sin(\xi +\lambda+\gamma)}{\sin(\xi)}, \qquad k_{-}(\lambda)=\frac{\sin(\xi -\lambda-\gamma)}{\sin(\xi)},
\en
where $\xi$ is the boundary parameter.

Using the above parametrization, the partition function of the six-vertex model with one reflecting end \cite{TSUCHIYA} can be written as,
\begin{align}
&Z_N(\{\lambda\},\{\mu\})=(\sin(2\gamma))^N \prod_{i=1}^N \sin(2 (\lambda_i+\gamma))\frac{\sin(\xi- \mu_i)}{\sin(\xi)} \nonumber \\
&\times \frac{\displaystyle\prod_{i,j=1}^N \sin(\gamma-(\lambda_i-\mu_j))\sin(\gamma+\lambda_i-\mu_j)\sin(\gamma-(\lambda_i+\mu_j))\sin(\gamma+\lambda_i+\mu_j)}{\displaystyle\prod_{\stackrel{i,j=1}{i<j}}^N -\sin(\lambda_j-\lambda_i)\sin(\mu_i-\mu_j) \sin(\lambda_j+\lambda_i)\sin(\mu_i+\mu_j)} \nonumber \\
&\times  \det{M},
\nonumber
\end{align}
where $M$ is a $N\times N$ matrix, whose matrix elements are $M_{ij}=\phi(\lambda_i,\mu_j)$ with
\eq
\phi(\lambda,\mu)=\frac{1}{\sin(\gamma-(\lambda-\mu))\sin(\gamma+\lambda -\mu)\sin(\gamma-(\lambda +\mu))\sin(\gamma+\lambda +\mu)}.
\nonumber
\en

\subsection{Homogeneous limit}

We can take the homogeneous limit along the same lines as \cite{KOREPIN1992}. This is done by taking $\lambda_i \rightarrow\lambda$ and $\mu_j \rightarrow\mu$. The main difference between the homogeneous limit of the six-vertex model with DWBC \cite{KOREPIN1992} and the present case, is that the partition function is no longer a function of the difference of the horizontal ($\{\lambda\}$) and vertical ($\{\mu\}$) spectral parameters. Therefore, in order to take these singular limits, we have to differentiate with respect to both variables. After a long but straightforward calculation we obtain,
\begin{align}
&Z_N(\lambda,\mu)=\left[\sin(2\gamma)\sin(2(\lambda+\gamma))\frac{\sin(\xi- \mu)}{\sin(\xi)}\right]^N \nonumber \\
&\times \frac{\displaystyle\left[\sin(\gamma-(\lambda-\mu))\sin(\gamma+\lambda-\mu)\sin(\gamma-(\lambda+\mu))\sin(\gamma+\lambda+\mu)\right]^{N^2}}{\displaystyle C_N \left[-\sin(2\lambda)\sin(2\mu)\right]^{\frac{N(N-1)}{2}} } \nonumber \\
&\times \tau_N(\lambda,\mu),
\label{homogeneous}
\end{align}
where $C_N=\left[\prod_{k=1}^{N-1} k!\right]^2$. The determinant is given by
\eq
\tau_N(\lambda,\mu)=\det(H),
\en
where the $H$-matrix elements are $H_{i,j}=(-\partial_{\mu})^{j-1}\partial_{\lambda}^{i-1}\phi(\lambda,\mu)$.

\subsection{Determinant and the bidimensional Toda equation}

The determinant expression $\tau_N(\lambda,\mu)$ is a bi-directional Wronskian solution of the bidimensional Toda equation \cite{DARBAUX,MA}, which reads
\eq
-\tau_N \partial_{\mu\lambda}^2\tau_N +(\partial_{\mu}\tau_N )(\partial_{\lambda}\tau_N ) =\tau_{N+1}  \tau_{N-1}.
\en
This equation can be conveniently written as
\eq
-\partial_{\mu\lambda}^2\left[\log(\tau_N)\right]=\frac{\tau_{N+1}\tau_{N-1}}{\tau_N^2}, \qquad N\geq 1,
\label{toda}
\en
which is supplemented by the initial data $\tau_0=1$ and $\tau_1=\phi(\lambda,\mu)$.

It is worth to note that this two variables determinant was firstly studied in \cite{DARBAUX} and later on \cite{MA}. In addition, the history of the single variable determinant, which is a solution of the Toda equation \cite{TODA,HIROTA}, goes back to \cite{SYLVESTER} and has interesting application to correlation functions in Ising model \cite{PERK}.

\subsection{Special solutions}

The partition function can be cast directly in simple expressions for some special points.

There is a special value of $\gamma$, where the partition function $Z_N(\lambda,\mu; \gamma)$ can be simply written as
\bear
Z_N(\lambda,\mu; \gamma=\frac{\pi}{4})=\left(\frac{\sin(\xi\mp\mu)}{\sin(\xi)}\right)^N  (\cos(2\lambda))^{\frac{N(N+1)}{2}}(\cos(2\mu))^{\frac{N(N-1)}{2}}.
\label{xypi4}
\ear

Additionally, for the cases where $\mu=\pm (\lambda+\gamma)$ and $\mu=\pm (\lambda-\gamma)$, the partition function is directly obtained
\begin{align}
&Z_N(\lambda,\pm\lambda\pm\gamma))= \label{zxx2} \\
&=\left(\frac{\sin(\xi\mp(\lambda+\gamma))}{\sin(\xi)}\right)^N (\sin(2\gamma))^{N^2} (-\sin(2\lambda))^{\frac{N(N-1)}{2}}(\sin(2(\lambda+\gamma)))^{\frac{N(N+1)}{2}}, \nonumber
\end{align}
\begin{align}
&Z_N(\lambda,\pm\lambda\mp\gamma)=\label{zxxga2}\\
&=\left(\frac{\sin(\xi\mp(\lambda-\gamma))\sin(2(\gamma+\lambda))}{\sin(\xi)} \right)^N (\sin(2\gamma))^{N^2} (\sin(2\lambda)\sin(2(\gamma-\lambda))^{\frac{N(N-1)}{2}}. \nonumber
\end{align}

We can easily take the thermodynamic limit of the above expressions, once it is clear that we have contribution of order $N^2$ with corrections of order $N$. The free energy  $F=-\lim_{N\rightarrow\infty}\frac{\log(Z_N)}{2N^2}$ (we set temperature to 1) is given by
\bear
e^{-2 F(\lambda,\mu;\gamma=\pi/4)}&=& \sqrt{\cos(2\lambda)\cos(2\mu)}, \label{sp1}\\
e^{-2 F(\lambda,\pm (\lambda+\gamma))}&=&\sin(2\gamma) \sqrt{-\sin(2\lambda)\sin(2(\lambda+\gamma))},\label{sp2} \\
e^{-2 F(\lambda,\pm (\lambda-\gamma))}&=&\sin(2\gamma) \sqrt{\sin(2\lambda)\sinh(2(\gamma-\lambda))}.\label{sp3}
\ear

We can also fix both spectral parameters $\lambda=\mu=0$ and anisotropy parameter $\gamma=\pi/3,\pi/4,\pi/6$. Using a more standard normalization where $a=b=1$, we obtain
\bear
Z_N(0,0 ; \frac{\pi}{3})=A_N^{VSASM}=\prod_{k=0}^{N-1}(3k+2)\frac{(6k+3)!(2k+1)!}{(4k+2)!(4k+3)!}=1,3,26,646,\dots
\label{vasm}
\ear
which is a combinatorial point connected to the number of vertically symmetric alternating sign matrices (VSASM). The relation between the six-vertex model with reflecting end and the VSASM was first noticed by Kuperberg \cite{KUPERBERG}. It is worth to note that when we fix $\mu=0$, the partition function (\ref{homogeneous}) becomes clearly independent of the boundary parameter $\xi$ even at finite $N$.

Other special cases are
\bear
Z_N(0,0 ;  \frac{\pi}{4})=2^N A_2^{VSASM}=2^{N^2},
\label{vasm2}
\ear
and
\bear
Z_N(0,0 ; \frac{\pi}{6})/3^N= A_3^{VSASM}= \frac{3^{N(N-3)/2}}{2^N} \prod_{k=1}^{N}\frac{(k-1)!(3k)!}{k((2k-1)!)^2}=1, 5, 126,\dots,
\label{vasm3}
\ear
where $A_x^{VSASM}$ are the $x$-enumeration of the vertically symmetric alternating sign matrices, in which a weight $x^k$ is given to each alternating sign matrix where $k$ is the number of $-1$ elements \cite{KUPERBERG}.

\section{Thermodynamic limit}\label{thermo}

We would like to consider the thermodynamic limit ($N\rightarrow \infty$) of the partition function (\ref{homogeneous}). For the case of large $N$, the partition function is expected to behave as 
\eq
Z_N(\lambda,\mu)=e^{-2 N^2 F(\lambda,\mu)+O(N)},
\en
where $F(\lambda,\mu)$ is the bulk free energy and we set temperature to 1.

In order to obtain the $F(\lambda,\mu)$, we proceed along the same lines as \cite{KOREPIN2000} and suppose the following ansatz for the large size behaviour of the determinant $\tau_N(\lambda,\mu)$,
\eq
\tau_N(\lambda,\mu)=C_N e^{2 N^2 f(\lambda,\mu)+O(N)},
\label{ansatz}
\en
where 
\eq
e^{-2 F(\lambda,\mu)}=
\frac{\sin(\gamma-(\lambda-\mu))\sin(\gamma+\lambda-\mu)\sin(\gamma-(\lambda+\mu))\sin(\gamma+\lambda+\mu)}{\sqrt{-\sin(2\lambda)\sin(2\mu)}}e^{2f(\lambda,\mu)}. 
\label{rel}
\en

Substituting the ansatz (\ref{ansatz}) in the Toda equation (\ref{toda}), we obtain the following differential equation for $f(\lambda,\mu)$,
\eq
-2\partial_{\mu\lambda}^2 f(\lambda,\mu) = e^{4 f(\lambda,\mu)},
\label{liouville-eq}
\en
which is the Liouville equation \cite{LIOUVILLE}. The general solution of this equation has the form of
\eq
e^{2 f(\lambda,\mu)}=\frac{\sqrt{-u^{'}(\lambda)v^{'}(\mu)}}{u(\lambda)+v(\mu)},
\label{sol}
\en
for arbitrary $C^2$ functions $u(\lambda),v(\mu)$ \cite{LIOUVILLE}.

In order to fix the function $f(\lambda,\mu)$ we need to impose boundary conditions on some meaningful solution (\ref{sol}) of the Liouville differential equation (\ref{liouville-eq}). The boundaries we have at our disposal are the exact solution of the partition function at special points described in the previous section. 

Our strategy is to chose $e^{2 f(\lambda,\mu)}$ to match with the solution at $\gamma=\frac{\pi}{4}$ (\ref{sp1}). This leave us a $\gamma$ dependent parameter to be determined. However the $\lambda,\mu$ dependence was already determined. In doing so, we obtain the following expression 
\eq
e^{2 f(\lambda,\mu)}=\frac{\alpha \sqrt{-\sin(\alpha\lambda)\sin(\alpha\mu)}}{\cos(\alpha\lambda)+\cos(\alpha\mu)}=\frac{\alpha \sqrt{-\sin(\alpha\lambda)\sin(\alpha\mu)}}{2\cos(\frac{\alpha}{2}(\lambda-\mu))\cos(\frac{\alpha}{2}(\lambda+\mu))},
\label{alphadep}
\en
where  $\alpha=\alpha(\gamma)$ is the undetermined parameter which is known only at $\alpha(\frac{\pi}{4})=4$.

We must use the boundary condition given by $\mu=\pm (\lambda+\gamma)$ (\ref{sp2}) to determine $\alpha$ parameter. Therefore we replace (\ref{alphadep}) on the  expression (\ref{rel}) and impose it to be equal to (\ref{sp2}). As a result, we immediately see that the only possible choice  for the parameter is $\alpha=\pi/\gamma$. The other points $\mu=\pm (\lambda-\gamma)$ are naturally fulfilled by this choice.

Therefore the free energy is completely determined as
\bear
e^{-2 F(\lambda,\mu)}&=&\frac{\sin(\gamma-\lambda+\mu)\sin(\gamma+\lambda-\mu)\sin(\gamma-\lambda-\mu)\sin(\gamma+\lambda+\mu)}{\sqrt{-\sin(2\lambda)\sin(2\mu)}} \nonumber \\
&\times &
\frac{\pi\sqrt{-\sin(\frac{\pi\lambda}{\gamma})\sin(\frac{\pi\mu}{\gamma})}}{2\gamma\cos(\frac{\pi(\lambda-\mu)}{2\gamma})\cos(\frac{\pi(\lambda+\mu)}{2\gamma})}.
\label{final-sol}
\ear

As an independent check, the solution obtained (\ref{final-sol}) at the special points $\gamma=\pi/3,\pi/4, \pi/6$ also coincides with the large-$N$ limit of the expressions (\ref{vasm}-\ref{vasm3}) \cite{KUPERBERG}.

\subsection{Ferrolectric phase: $\Delta>1$}

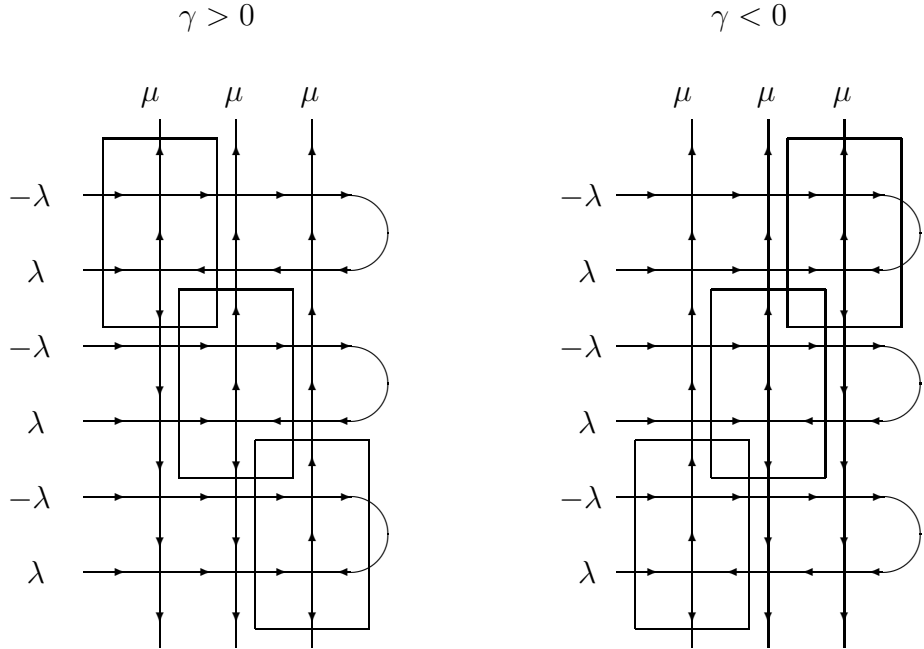
\begin{figure}[h]
\unitlength=0.5mm
\begin{center}
\begin{picture}(200,165)(-20,-10)

\put(5,145){$\gamma>0$}

\multiput(-20,0)(0,20){6}{\line(1,0){70}}
\multiput(0,-20)(20,0){3}{\line(0,1){140}}
\multiput(50,10)(0,40){3}{\oval(20, 20)[r]}
\multiput(-10,0)(0,20){6}{\vector(1,0){1}}
\multiput(0,-12.5)(20,0){3}{\vector(0,-1){1}}
\multiput(0,112.5)(20,0){3}{\vector(0,1){1}}
\multiput(47.5,0)(0,40){3}{\vector(-1,0){1}}
\multiput(49.5,20)(0,40){3}{\vector(1,0){1}}
\multiput(12.5,0)(20,0){2}{\vector(1,0){1}}
\multiput(12.5,20)(20,0){2}{\vector(1,0){1}}
\multiput(12.5,40)(20,0){1}{\vector(1,0){1}}
\multiput(30,40)(20,0){1}{\vector(-1,0){1}}
\multiput(12.5,60)(20,0){2}{\vector(1,0){1}}
\multiput(10,80)(20,0){2}{\vector(-1,0){1}}
\multiput(12.5,100)(20,0){2}{\vector(1,0){1}}
\multiput(0,7.5)(20,0){2}{\vector(0,-1){1}}
\multiput(40,10)(20,0){1}{\vector(0,1){1}}
\multiput(0,27.5)(20,0){2}{\vector(0,-1){1}}
\multiput(40,30)(20,0){1}{\vector(0,1){1}}
\multiput(0,47.5)(20,0){1}{\vector(0,-1){1}}
\multiput(20,50)(20,0){2}{\vector(0,1){1}}
\multiput(0,67.5)(20,0){1}{\vector(0,-1){1}}
\multiput(20,70)(20,0){2}{\vector(0,1){1}}
\multiput(0,90)(20,0){3}{\vector(0,1){1}}

\put(-15,65){\line(0,1){50}}
\put(15,65){\line(0,1){50}}
\put(-15,65){\line(1,0){30}}
\put(-15,115){\line(1,0){30}}

\put(5,25){\line(0,1){50}}
\put(35,25){\line(0,1){50}}
\put(5,25){\line(1,0){30}}
\put(5,75){\line(1,0){30}}

\put(25,-15){\line(0,1){50}}
\put(55,-15){\line(0,1){50}}
\put(25,-15){\line(1,0){30}}
\put(25,35){\line(1,0){30}}

\put(-35,-3){$\lambda$}
\put(-40,17){$-\lambda$}
\put(-35,37){$\lambda$}
\put(-40,57){$-\lambda$}
\put(-35,77){$\lambda$}
\put(-40,97){$-\lambda$}

\put(-5,125){$\mu$}
\put(17,125){$\mu$}
\put(37,125){$\mu$}


\put(145,145){$\gamma<0$}

\multiput(120,0)(0,20){6}{\line(1,0){70}}
\multiput(140,-20)(20,0){3}{\line(0,1){140}}
\multiput(190,10)(0,40){3}{\oval(20, 20)[r]}
\multiput(130,0)(0,20){6}{\vector(1,0){1}}
\multiput(140,-12.5)(20,0){3}{\vector(0,-1){1}}
\multiput(140,112.5)(20,0){3}{\vector(0,1){1}}
\multiput(187.5,0)(0,40){3}{\vector(-1,0){1}}
\multiput(189.5,20)(0,40){3}{\vector(1,0){1}}
\multiput(150,0)(20,0){2}{\vector(-1,0){1}}
\multiput(152.5,20)(20,0){2}{\vector(1,0){1}}
\multiput(152.5,40)(20,0){1}{\vector(1,0){1}}
\multiput(170,40)(20,0){1}{\vector(-1,0){1}}
\multiput(152.5,60)(20,0){2}{\vector(1,0){1}}
\multiput(152.5,80)(20,0){2}{\vector(1,0){1}}
\multiput(152.5,100)(20,0){2}{\vector(1,0){1}}
\multiput(140,10)(20,0){1}{\vector(0,1){1}}
\multiput(160,7.5)(20,0){2}{\vector(0,-1){1}}
\multiput(140,30)(20,0){1}{\vector(0,1){1}}
\multiput(160,27.5)(20,0){2}{\vector(0,-1){1}}
\multiput(140,50)(20,0){2}{\vector(0,1){1}}
\multiput(180,47.5)(20,0){1}{\vector(0,-1){1}}
\multiput(140,70)(20,0){2}{\vector(0,1){1}}
\multiput(180,67.5)(20,0){1}{\vector(0,-1){1}}
\multiput(140,90)(20,0){3}{\vector(0,1){1}}

\put(165,65){\line(0,1){50}}
\put(195,65){\line(0,1){50}}
\put(165,65){\line(1,0){30}}
\put(165,115){\line(1,0){30}}

\put(145,25){\line(0,1){50}}
\put(175,25){\line(0,1){50}}
\put(145,25){\line(1,0){30}}
\put(145,75){\line(1,0){30}}

\put(125,-15){\line(0,1){50}}
\put(155,-15){\line(0,1){50}}
\put(125,-15){\line(1,0){30}}
\put(125,35){\line(1,0){30}}

\put(110,-3){$\lambda$}
\put(105,17){$-\lambda$}
\put(110,37){$\lambda$}
\put(105,57){$-\lambda$}
\put(110,77){$\lambda$}
\put(105,97){$-\lambda$}

\put(135,125){$\mu$}
\put(157,125){$\mu$}
\put(177,125){$\mu$}

\end{picture}
\end{center}
\caption{The dominant state in the regime $\Delta>1$ for $\mu>0$. The case $\mu<0$ is obtained by mirror image. The boxes are a guide to indicate a pattern along the diagonal which segregates the vertex configurations above and below the diagonal.}
\label{ferro}
\end{figure}

In the case $\Delta>1$, one can obtain the expression for the free energy in the thermodynamic limit looking at the leading order state (see Figure \ref{ferro}), analogously to the case of DWBC\cite{KOREPIN2000}. The expression for the free energy can be written as
\eq
e^{-2 F(\lambda,\mu)}=\sinh(\lambda-|\mu|+|\gamma|)\sqrt{\sinh(\lambda+|\mu|-\gamma)\sinh(\lambda+|\mu|+\gamma)},
\label{final-ferro}
\en
where we have used the following parametrization for the Boltzmann weights
\eq
a(\lambda)=\sinh(\lambda-\gamma),\qquad b(\lambda)=\sinh(\lambda+\gamma)
, \qquad c(\lambda)=\sinh(2|\gamma|),
\en
which implies $\Delta=\cosh(2\gamma)$.

However due to the lack of suitable boundary condition, we are unable to fix the solution of Liouville equation that matches with the expected formula (\ref{final-ferro}). A precise determination of the Liouville solution for this case has elude us so far.

\section{Entropy}\label{ent}

The entropy as a function of temperature can be obtained from the free energy expression and it differs from the case of domain wall boundary conditions at finite temperatures. However, it is worth to note that the entropy of the six-vertex model with reflecting end is exactly same as the entropy of the six-vertex model with domain wall boundary at infinite temperature. 

One can compute the infinite temperature entropy directly from the free energy expression. This is obtained by tuning the Boltzmann weights to be all equal $a_{\pm}=b_{\pm}=c=1$. 

First, we set $a_{\pm}=b_{\pm}=1$ by fixing $\lambda=\mu=0$ and assuming suitable normalization, this implies that (\ref{final-sol}) reads
\eq
e^{-F(0,0;\gamma)}=\frac{\pi}{2}\frac{\sin(\gamma)}{\gamma}.
\label{infty}
\en
which agrees with the case of domain wall boundary for any $\gamma$ value \cite{KOREPIN2000,ZINNJUSTIN}.

The entropy per lattice site is directly obtained from (\ref{infty}) for $\gamma=\pi/3$, which results
\eq
S=\frac{1}{2}\llg{\left(\frac{3^3}{2^4}\right)}.
\label{ent-tsuchiya}
\en
Naturally, this value can also be obtained by taking the large-$N$ limit of (\ref{vasm}), which coincides with the number of vertically symmetric alternating sign matrices ($A_N^{VSASM}$)\cite{ROBBINS}. 

Analogously, one can compute the entropy from the large-$N$ limit of the partition function with DWBC at the point where $a=b=c=1$. The partition function at this point is given by \cite{KUPERBERG1996,KOREPIN2000},
\eq
Z_N^{DWBC}(\lambda-\mu=\frac{\pi}{3};\gamma= \frac{\pi}{3})=A_N^{ASM}=\prod_{k=0}^{N-1} \frac{(3k+1)!}{(N+k)!}=1,2,7,42,429,\cdots,
\label{asm}
\en
which coincides with the number of alternating sign matrices ($A_N^{ASM}$)\cite{ASM}. Therefore, the entropy is given by,
\eq
S_{DWBC}=\lim_{N\rightarrow \infty}\frac{1}{2 N+1}\llg\left(\frac{A_{N+1}^{ASM}}{A_N^{ASM}}\right)=\frac{1}{2} \ln\left(\frac{3^3}{2^4}\right).
\label{ent-dwbc}
\en
which shows that the agreement between $S$ and $S_{DWBC}$, although the expressions (\ref{vasm}) and (\ref{asm}) are different at finite-$N$.

This agreement can be simply understood in the context of the alternating sign matrices. At infinite temperature, the partition function (\ref{vasm}) is roughly just counting the number of equally likely physical states, which coincides with the number of vertically symmetric alternating sign matrix ($A_N^{VSASM}$). Likewise for the case of the partition function with DWBC (\ref{asm}), which is equal to the number of alternating sign matrices ($A_N^{ASM}$). In particular, one has that any vertically symmetric alternating sign matrix is an alternating sign matrix, since they are a special subset of the alternating sign matrices\cite{RAZUMOV}. In other words, the following relation $A_N^{VSASM} \sim (A_N^{ASM})^2$ holds for large $N$. Taking in account that the Tsuchiya partition function describes the six-vertex model on a $N \times 2 N$ lattice, that is, twice bigger than the domain wall lattice $N \times N$, one sees that both entropies coincide. 

Similarly, one has the same large $N$ relation among the $x$-enumeration expressions (\ref{vasm2}-\ref{vasm3}) and its counterparts for alternating sign matrix \cite{KUPERBERG}. This gives some explanation for the agreement between free-energy of the six-vertex model with reflecting end 
and domain wall boundary on the line $a_{\pm}=b_{\pm}$, which holds for arbitrary $\gamma$ values.

\section{Conclusion}
\label{CONCLUSION}

In this paper we computed the free energy in the thermodynamic limit of the six-vertex model with domain wall and reflecting end in the disordered regime $-1<\Delta<1$. The homogeneous limit of the Tsuchiya partition function formula was discussed. Using the fact that the determinant formula in the homogeneous limit is a solution of the bidimensional Toda equation, we showed that the function which control the large-$N$ limit of the partition function is a solution of the Liouville partial differential equation. We were able to find a suitable solution of this differential equation in the disordered regime. 

We have also computed the entropy at infinite temperature. We noted that at infinite temperature the entropy of the six-vertex with reflecting end coincides with the entropy of the six-vertex model with domain wall boundary conditions. However, as it is largely known \cite{KOREPIN2000}, this value is different from the case of periodic boundary condition. This is another example where the physical properties in the infinite size limit depend on the boundary choice. One could rise the question about the existence of spatial phase separation in the case of reflecting end boundary and what would the the analogue of the artic circle \cite{PRONKO}.

An explicit formula for the free energy in the ferroelectric regime $\Delta>1$ was given based on the leading order state. The dominant state resembles the dominant ferroelectric state in the case of domain wall boundary condition. In that case, there is a separation line along the diagonal segregating different vertices. However due to the lack of suitable boundary conditions, we were unable to fix a solution of Liouville equation which agrees with our formula. We intend to address the other phases in the future.

Finally, we would like to remark that it would be interesting to consider the case of non-diagonal boundary \cite{NONDIAGONAL}, where additional configuration might be allowed due to the boundary.

\section*{Acknowledgments}
V.E. Korepin was supported by NSF Grant DMS 1205422. G.A.P. Ribeiro acknowledges financial support through the grant 2012/24514-0, S\~ao Paulo Research Foundation (FAPESP).


\begin{thebibliography}{100}
\bibitem{BAXTER} R.J. Baxter \textit{Exactly solved models in statistical mechanics} (AP, London, 1982).
\bibitem{BOOK} V.E. Korepin, N.M. Bogoliubov, and A.G. Izergin \textit{Quantum inverse scattering method and correlation functions} (CUP, Cambridge, 1993).
\bibitem{FADDEEV} L.A. Takhtajan and L.D. Faddeev, {\em Russian Math. Surveys, 34 (1979) 11}
\bibitem{KOREPIN1982} V.E. Korepin, Comm. Math. Phys., 86 (1982) 361.
\bibitem{KOREPIN1992} A.G. Izergin, D.A. Coker and V.E. Korepin, J. Phys. A: Math. Gen. 25 (1992) 4315.
\bibitem{KOREPIN2000} V.E. Korepin and P. Zinn-Justin, J. Phys. A 33 (2000) 7053.
\bibitem{ZINNJUSTIN} P. Zinn-Justin, Phys. Rev. E 62 (2000) 3411.
\bibitem{BLEHER} P.M. Bleher, V.V. Fokin, Comm. Math. Phys., 268 (2006) 223; P.M. Bleher, K. Liechty, Comm. Math. Phys. 286 (2009) 777; P.M. Bleher, K. Liechty, J. Stat. Phys. 134 (2009) 463; P.M. Bleher, K. Liechty, Comm. on Pure and Appl. Math., 63 (2010) 779.
\bibitem{SKYLIANIN} I. Cherednik, Theor. Math. Phys. 61 (1984) 977; E.K. Sklyanin, J. Phys. A: Math. Gen. 21 (1988) 2375.
\bibitem{TSUCHIYA} O. Tsuchiya, J. Math. Phys., 39 (1998) 5946.
\bibitem{KUPERBERG} G. Kuperberg, Ann. of Math. 156 (2002) 835.
\bibitem{GALLEAS} W. Galleas, J. Lamers, Nucl. Phys. B, 886 (2014) 1003.
\bibitem{COLOMO} F. Colomo and A.G. Pronko, Nucl. Phys B, 798 (2008) 340; N.M. Bogoliubov, A.G. Pronko and M.B. Zvonarev, J. Phys. A: Math. Gen. 35 (2002) 5525.
\bibitem{MOTEGI} K. Motegi, Physica A 390 (2011) 3337; K. Motegi, Rep. Math. Phys. 67 (2011) 87.
\bibitem{GOHMANN} F. G\"ohmann, M. Bortz, H. Frahm, J. Phys. A: Math. Gen. 38 (2005) 10879; K.K. Kozlowski and B. Pozsgay, J. Stat. Mech. (2012) P05021.
\bibitem{MA} Wen-Xiu Ma, Phys. Lett. A 375 (2011) 3931.
\bibitem{RAZUMOV} A.V. Razumov and Yu. G. Stroganov, Theor. Math. Phys. 141 (2004) 1609.
\bibitem{KUPERBERG1996} G. Kuperberg, Int. Math. Res. Notices 3 (1996) 139.
\bibitem{LIOUVILLE} J. Liouville, J. Math. 18 (1853) 71; D.G. Crowdy, Int. J. Engng Sci. 35 (1997) 141.
\bibitem{ROBBINS} D.P. Robbins, Math. Intel. 13 (1991) 12.
\bibitem{ASM} G. Andrews, J. Combin. Theor. Ser. A 66 (1994)28; D. Zeilberger, Elect. J. Combin. 3 (1996) R13: 1-84.
\bibitem{PRONKO} F. Colomo and A.G. Pronko, J. Phys. A 37 (2004) 1987; J. Stat. Phys., 138 (2010) 662.
\bibitem{NONDIAGONAL} Wen-Li Yang et al., Nucl. Phys. B 847 (2011) 367; Wen-Li Yang et al., Nucl. Phys. B 844 (2011) 289.
\bibitem{DARBAUX} G. Darboux, Le\c{c}ons sur la Th\'eorie G\'en\'erales des Surfaces et les Applications G\'eom\'etriques du Calcul Infinit\'esimal, Deuxi\`eme Partie (Gauthiers-Villars et fils, Paris, 1889), Livre IV, Ch. VI, p. 124.
\bibitem{SYLVESTER} J.J. Sylvester, Compt. Rend. Acad. Sc. 54 (1862) 129.
\bibitem{TODA} M. Toda, J. Phys. Soc. Japan, 22 (1967) 431.
\bibitem{HIROTA} R. Hirota, J. Phys. Soc. Japan, 35 (1973) 286.
\bibitem{PERK} H. Au-Yang and J.H.H. Perk, Phys. D, 18 (1986) 365; H. Au-Yang and J.H.H. Perk, Phys. A, 144 (1987) 44.
\end{thebibliography}
\end{document}